\documentclass[prb,preprint]{revtex4-1} 

\usepackage{amsmath} 

\usepackage{graphicx} 
\usepackage{epstopdf}
\usepackage{slashed}
\usepackage{hyperref}
\usepackage{url}

\begin{document}
\raggedbottom

\title{Special Relativity from the Dynamical Viewpoint}

\author{William M. Nelson}
\email{wmn@cox.net}

\received{May 14, 2014}
\accepted{}

\begin{abstract}
Arguments are reviewed and extended in favor of presenting special relativity at least in part from 
a more mechanistic point of view.  A number of generic mechanisms are
catalogued and illustrated with the goal of making relativistic effects
seem more natural by connecting with more elementary aspects of physics,
particularly the physics of waves. 

 \end{abstract}
\maketitle

\section{\label{sec:intro}Introduction}

Qualitative understanding in physics usually comes from the
same calculations which give quantitative knowledge, but in the case of special relativity 
this connection is weakened. Lorentz invariance enables computation of many results
 with little or no consideration of microscopic details, hence
there is usually no practical benefit (and often great difficulty) in analyzing
a given scenario in more detail. 

Complete reliance on Lorentz invariance can, nevertheless, leave
the uncomfortable sensation of a gap in understanding.  Generally one expects
concrete effects to have concrete causes, yet in relativity one finds
very concrete-sounding effects (slowing clocks, shortening objects) whose cause
is variously attributed to an abstract principle (invariance of the speed of light) 
or to an abstract entity (spacetime) or
to conventions (how things are measured, how simultaneity is defined). 
We don't claim that these explanations are incorrect but it seems
reasonable to look also for more ordinary physical explanations and try to 
understand how they connect to the abstract notions. 

Adding to the sense of abstraction in special relativity 
 is the lack of straightforward experimental
demonstration for some of the central phenomena. Time dilation
has long been observed directly in experiments using moving clocks,\cite{planes, opticalclocks}
and mass/energy equivalence has been similarly verified,\cite{directtest}
but length contraction and the synchronization discrepancy for separated clocks
are more challenging. For these effects one must appeal to 
indirect observations (e.g., the Michelson-Morley negative result)
and the overall consistency of the framework. 

The ``dynamical'' approach to relativity aims to 
shrink these gaps by tracing relativistic effects to their underlying physical mechanisms, 
at least qualitatively. This helps to flesh out the abstract arguments 
and makes the hard-to-demonstrate effects more believable by showing that they 
have straightforward causes rooted in familiar physics. 

We emphasize that seeking mechanistic accounts for relativistic effects does not
mean postulating a preferred reference frame or medium.
The point is rather 
that one can, in principle, compute any relativistic quantity (e.g., the size of a moving
molecule) directly from the underlying theories of matter 
without invoking relativity at all.
In practice this is very difficult but a
mechanistic analysis still helps to show qualitatively how
the effects arise.

The mechanisms most relevant to relativity are those of waves and fields. Relativity 
developed simultaneously with electromagnetism, the first fundamental theory based on fields and waves,  
and this paradigm now extends to all known matter (in the ``standard model'' of particle physics). 
Many relativistic phenomena that at first seem strange or opaque become 
quite natural when viewed in a field/wave context. 

It is worth noting that many elements of the dynamical viewpoint pre-date relativity.
Indeed, many relativistic phenomena were first discovered through 
dynamical considerations, starting over two decades before Einstein's work.
FitzGerald, Lorentz,
Larmor, and Thompson were all motivated by dynamics in introducing 
the seminal notions of length contraction, time dilation,
and relativistic mass.\cite{brown2,brown3,larmor,thomp} 

Following Einstein and Minkowski the dynamical view languished, displaced
by the principle-based treatment that was more efficient and also did not require a microscopic
understanding of matter, which was not available at that time. 
The viewpoint was revived by J.S. Bell in his 1976 essay ``How to teach 
special relativity,'' \cite{bell} 
which, however, had little impact on pedagogy, possibly because it 
proposed rather opaque numerical computations. In the 1990's the baton
was picked up by H.R. Brown, often in collaboration with O. Pooley,
who mounted a vigorous philosophical defense of Bell's viewpoint and extended it to
cover general relativity as well.\cite{brown4,brown5,brown,dynamic}  
Fully constructive examples have been presented by D.J. Miller,
who also makes the suggestion, correct in our view, that the primary 
aim of the dynamical approach should be to supplement the customary one
with increased qualitative understanding.\cite{miller}  
The dynamical viewpoint has also been presented to laypeople, somewhat briefly by N. D. Mermin,\cite{mermin} 
and more fully by the present author.\cite{rmr}

The main aim in what follows is to catalog and illustrate some of the principle mechanisms that underlie
relativistic phenomena.  Our focus is narrower than that of
Brown and Pooley in that we do not discuss general relativity
nor (in any detail) the historical development of the theories; also, we have tried to
avoid taking positions on philosophical questions such as the primacy of spacetime. 
Our approach differs from that of Miller (and Bell) in that we do not attempt 
a full constructive derivation but instead emphasize qualitative behavior.

The manuscript is organized as follows.  Section~\ref{sec:motive} describes in more detail what the dynamical view means, 
at least in the approach taken here.
Section~\ref{sec:mechanism} enumerates mechanisms that give rise to relativistic effects
and shows models to demonstrate them. 
Section~\ref{sec:frames} discusses further how mechanistic explanations connect to the
more customary approach, and Sec.~\ref{sec:conc} provides a brief conclusion.

\section{\label{sec:motive}Meaning and Context of the Dynamical Viewpoint}

The dynamical viewpoint aims to connect the phenomena of relativity 
to underlying physical aspects of the universe as currently understood.
The phrase ``underlying physical aspects'' could be interpreted many ways, 
but we mean here the generic characteristics of   
 current state-of-the-art theories, namely field theories, 
{\it excluding} Lorentz invariance. 
  
In the most simplistic (but still useful) formulation this means starting 
with a stipulation that everything in the 
world is ``made from waves.''\cite{shupe} Particles are really wave packets, and
composite objects consist of wave packets moving under the influence of other fields
whose effects are also transmitted by propagating waves. 
The main goal is then to translate generic wave and field knowledge
 into intuition about relativistic phenomena. 

Taking a dynamical view it is natural to think not just about Lorentz-invariant
theories but also about related theories that share the
same dynamical mechanisms seen in our universe.  
 A generic change in the parameters of a Lorentz-invariant field theory leads to 
a theory that is still a field theory, hence shares the same types of 
mechanisms and the same ``relativity-like'' effects, but which is no 
longer precisely Lorentz-invariant. For example, one might alter the
wave speeds of the different fields to be direction-dependent and/or unequal to each other. 
Considering this wider context of theories helps to illuminate Lorentz 
invariance by contrast, much as one understands rotation invariance 
in mechanics by considering both symmetric and non-symmetric potentials. 
 
We certainly do not wish to suggest that either Lorentz invariance or Minkowski space are not fundamental; 
the point is merely to provide a broader
picture of where these concepts come from and why the prior Newtonian picture, in which motion {\it per se} 
entails no real effects, is not compatible with a world consisting of fundamental 
fields. 

\section{\label{sec:mechanism}Mechanisms}

We describe some of the main mechanisms giving rise to relativistic effects and 
show elementary models that illustrate them. Although the models do sometimes  
 produce quantitatively correct relativistic answers, 
the intent is never to independently prove Lorentz invariance 
but rather to illustrate generic mechanisms that create the {\it possibility} 
of Lorentz invariance in a world composed of fields and waves. 

For the moment we take a naive view of concepts such as motion, reference frame,
and measurement, defering a deeper discussion to Sec.~\ref{sec:frames}. 
We will sometimes refer to relatively-moving observers as ``moving'' and ``stationary,'' however,
these are meant merely as convenient labels.


\subsection{Wave propagation and rigidity}
\label{WPR}

One immediate consequence of the field/wave paradigm is non-rigidity of objects. 
Rigid objects can exist in Newtonian theories because forces
propagate at infinite speed, but in a field theory universe all
forces must propagate via waves, and waves inherently move at finite speed. 
For everyday objects in our world, which are built from atoms, the forces 
binding them are of course mainly electrical and the waves electromagnetic.  

But finite speed of force propagation means that all objects will deform under an applied force, simply
because one part starts to move before the other parts experience any force at all.
This fundamental non-rigidity is independent of the strength or organization of bonds
within the object. 

This certainly does not prove Lorentz contraction, and additional study is needed 
to see whether a deformation persists 
after the acceleration stops (see Sec.~\ref{LCDPF} below).
Nevertheless, the failure of rigidity does at least create the possibility 
of contraction.\cite{rigidity}


\subsection{Wave propagation and time dilation}
\label{WPTD}

Not only do waves propagate at finite speed but wave speed is also 
generically independent of the motion of the source. The speed of sound waves from a jet doesn't depend
on the speed of the jet; the speed of water waves in a wake doesn't depend on the speed of the boat.
We stress that this is
{\it source-independence}, not {\it observer-independence}, the vastly stronger
assertion that implies complete Lorentz invariance. Source-independence is a generic feature of
wave physics, while observer-idependence is an assertion not about the waves themselves but about
complex physical effects occurring within every possible apparatus
 that could be used to measure wave speed.   

\begin{figure}[h!]
\includegraphics[width=8.5cm]{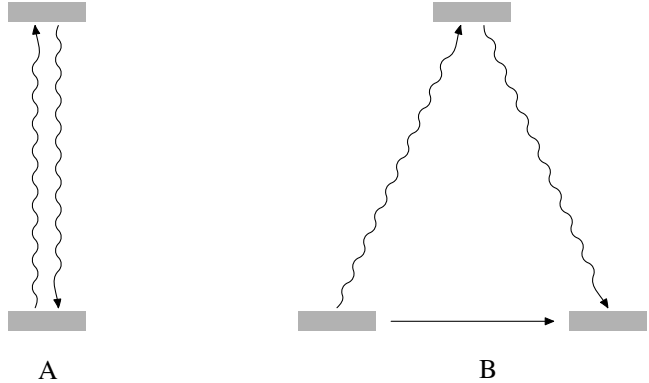}
\caption{Stationary (left) and moving (right) light clocks. Motion is to the right; the bottom mirror is drawn twice to show motion.}
\label{fig:lightclock}
\end{figure}

Source-independence of wave speeds causes changes to the tick rates of moving clocks, 
as demonstrated most clearly within the venerable ``light clock.'' The light clock consists of
a light pulse bouncing back and forth between two mirrors, as shown in Fig.~\ref{fig:lightclock}\,(a).

If we now imagine the clock placed on board a spaceship with glass walls and flown by
us at high speed, it will look as in Fig.~\ref{fig:lightclock}\,(b) (taking the clock to be oriented 
transverse to the motion of the ship). The light pulse now travels
a longer distance for each cycle, hence the tick rate is slower. Source-independence 
prevents the clock from making any kind of automatic adjustment to preserve its rate
when moving; source motion alters the spatial pattern of the waves (Doppler effect) but this 
does not help the clock to maintain its rate. 
All clocks will be affected by this effect to some degree because their 
subcomponents can only interact through wave transmission.

In this case a simple calculation using the Pythagorean theorem does
give the correct Lorentz time dilation factor.\cite{lccalc} 
It should, however, be recognized
that this calculation relies on additional implicit assumptions. 
Indeed, Fig.~\ref{fig:lightclock} could also be drawn within a non-Lorentz-invariant 
theory having, for example, different wave speeds in different directions, and the naive 
calculation would then be wrong. What remains generically true, however, 
is that the moving clock will change its rate due to source-independence of wave speed. 

Another possibility is that shape changes as discussed above
(and in Sec.~\ref{LCDPF} below) could counteract the rate change due to wave propagation. 
If the light clock housing were to shrink, thus reducing the vertical travel distance, 
then the rate could remain unchanged; however, there is
no reason to expect the effects to conspire in this way.\cite{conspiracy}


\subsection{Wave packets, group velocity, and relativistic mass}
\label{WPGVRM}

The relativistic mass increase
\begin{equation}
\label{relmass}
m_{\rm rel} = \gamma m,
\end{equation}
where $\gamma \equiv \left(1-v^2/c^2\right)^{-1/2}$,
seems  ad hoc when introduced in the context of particle mechanics.
It is difficult to understand why an elementary and indivisible piece of matter should become
harder to accelerate when moving faster. 
The effect is, however, quite natural for particles viewed as 
 wave packets, as in modern quantum theories. 
Close analogies to such ``matter waves'' can be constructed using 
simple wave-on-string models, allowing the effect to be understood
in a simple setting.

For simplicity we start by considering matter moving only in the $x$-direction. 
We then suppose that the matter is actually
described by a wave $\xi$ obeying the standard wave equation
for transverse waves on a string:
\begin{align}
\label{m=0}
-\partial_t^2\xi + \partial_x^2\xi = 0,
\end{align}
where we have set all the constants to one for simplicity. 

All traveling waves in this model move with fixed speed $c=1$ regardless
of frequency, hence the model corresponds to ``massless'' waves 
such as light. To model massive particles
one needs an extra restoring force at each point. This can be visualized
as placing Hooke's law springs on either side of the
the string at frequent intervals, hence we will refer to the model as
``$\mbox{\rm string}+\mbox{\rm springs}$.''
The equation we will use is the continuous limit in which the restoring
force acts at each point:
\begin{align}
\label{m>0}
-\partial_t^2\xi + \partial_x^2\xi - m^2\xi = 0.
\end{align}
Here the restoring force constant is labeled as $m^2$ anticipating that $m$ will be
 the mass of a ``particle'' in this model. 

The massive case with $m > 0$ differs qualitatively from the massless case ($m=0$) in two important ways. First, the
presence of the springs obstructs the waves and slows them down (e.g., a single spring attached
to a string will reflect some fraction of incident waves). 
Second, the massive waves can approximately sit still, because they can 
oscillate in place under the spring restoring force. The massless waves
cannot sit still because their only restoring force comes from neighboring parts of the string.  

The elementary traveling wave solutions take the usual form of
$\sin(kx \pm \omega t)$ and $\cos(kx\pm \omega t)$,
where the angular velocity $\omega$ and wavenumber $k$ satisfy
\begin{align}
\label{omegak}
\omega^2 = k^2 + m^2.
\end{align}
These solutions extend over all space and don't look much like particles, 
but this can be remedied by building a ``wave packet''---a
superposition of waves having wavenumbers in a narrow range, such as\cite{lonngren}
\begin{align}
\label{packet}
\xi(x,t) = \int_{\bar{k}-\Delta k}^{\bar{k}+\Delta k} dk \,\cos(kx - \omega t).
\end{align}
Looking first at $t=0$, we see that the waves are all in phase at $x=0$, but they interfere increasingly destructively 
away from this point.
This creates 
a localized packet having approximate location $\bar{x}=0$ and approximate wavenumber $\bar{k}$
(and corresponding angular velocity $\bar{\omega}$). 
As $t$ changes, the location of the in-phase maximum moves, and with it the wave packet. By substituting
\begin{align}
\omega \approx \bar{\omega} + {\partial\omega\over\partial k}\bigg{|}_{\bar{k}}\left(k-\bar{k}\right) 
\end{align}
into Eq.~(\ref{packet}), and using Eq.~(\ref{omegak}), one sees that the packet moves with approximate velocity given by the ``group velocity''
\begin{align}
\label{gv}
v = {\partial\omega\over\partial k}\bigg|_{\bar{k}} = {\bar{k}\over\bar{\omega}}.
\end{align}
We henceforth drop the bar notation and use $k$ and $\omega$ for the packet's central values. 

The $\mbox{\rm string}+\mbox{\rm spring}$ model is not so far from the real description of particles in 
modern quantum theories. The non-particle-like extended solutions do exist in nature, 
but they are converted to more localized wave packets
through interaction with other clumps of matter (such as measuring devices).\cite{mott} 
Quantization also crucially prevents the waves from dissipating away to zero. 

A wave packet, like a particle, can be accelerated by
a potential field. For example, letting the potential be $\phi(x)$ one
can add a coupling term $\phi(x)\xi$ to Eq.~(\ref{m>0}); this corresponds to 
letting the ``spring tension'' $m^2$ vary with position, which is essentially the 
action of the standard model Higgs field.\cite{coupling} 

The qualitative origin of relativistic mass can be seen immediately 
since the (absolute value of) group velocity satisfies $v < 1$ for 
all values of $k$. The packets can never attain the ``spring-free'' speed $v = 1$ 
because their propagation is
hindered by the springs.  As the limiting speed is approached, energy applied to accelerate
a packet goes instead into vibrations of the field.
Indeed the energy {\it always} goes into field vibrations and the packet
acceleration is merely a side effect that occurs for low speeds.

To see this in more detail we start with the
standard expressions for energy and momentum of a vibrating string, with the
Hooke's law energy added:\cite{lonngren}
\begin{align}
\label{E}
\mathcal{E} &= \int dx\, {1\over 2}\left[(\dot{\xi})^2 + (\xi^\prime)^2 + m^2\xi^2\right] \\
\mathcal{P} &= -\int dx\, \dot{\xi}\xi^\prime.
\label{P}
\end{align}
Here, the dot and prime indicate derivatives with respect to time and space.  Evaluating these for a wave packet that is narrow in $k$ space gives to good accuracy
\begin{align}
\label{Ebar}
E &= \omega^2N^2 \\
P &= \omega kN^2
\label{Pbar}
\end{align}
where $N^2$ is the squared norm:\cite{packetnorm}  $N^2 = \int dx\, \xi^2$.

To go further in the program of constructing particles out of wave packets 
one has to decide what value of $N^2$ constitutes a single particle. 
Not just any arbitrary convention will do, but it should be 
preserved, at a minimum, under slowly-varying (``adiabatic'') conditions. 
Under adiabatic conditions the potential fields vary weakly in space and time,
creating only small forces that change slowly.  
A single particle moving in such a weak field should remain as a single 
particle, although its amplitude $N^2$ may change. 

Similar questions were studied in the early days of quantum mechanics and it
was shown that certain quantities are invariant under
 adiabatic changes.
The most well-known occurs in the harmonic oscillator and
 takes the form $E/\omega$.\cite{siklos}
This invariant also applies to wave packets because the vibrating string is just
a collection of harmonic oscillators, one for each $k$, as can be seen by Fourier-transforming
Eq.~(\ref{m>0}) in the spatial variable. 

Making use of Eq.~\eqref{Ebar}, we see that the wavepacket norm $N$ will evolve such that
\begin{align}
\label{Nomega1}
{E\over\omega} = \omega N^2 = \mbox{\rm const.}, 
\end{align}
and the single-particle normalization definition should be consistent with this. 
We choose the simplest option,
\begin{align}
\label{Nomega}
N(\omega) = {1\over \sqrt{\omega}},
\end{align}
which is also the normalization arrived at through quantization.
Applying Eq.~(\ref{Nomega}) to Eqs.~(\ref{Ebar}) and \eqref{Pbar}, one finds for the single-particle energy and momentum
\begin{align}
\label{Ebar2}
E &= \omega \\
P &= k,
\label{Pbar2}
\end{align}
which are the well-known relations proposed by Einstein for photons and by
de Broglie for matter waves (in units with $\hbar = c =1$).

A force $F$ arising from some potential $\phi(x)$ 
by definition acts to change the wave packet momentum by
\begin{align}
\label{F}
\dot{P} = F,
\end{align}
and hence from Eq.~(\ref{Pbar2})
\begin{align}
\label{F2}
F = \dot{k},
\end{align}
which is exactly as expected for a force acting on the relativistic mass Eq.~(\ref{relmass}),
since Eqs.~\eqref{omegak} and \eqref{gv} imply  
\begin{align}
\label{omegagamma}
\omega = m\gamma
\end{align}
and 
\begin{align}
\label{omegagamma2}
k = m\gamma v.
\end{align}
Thus, the relativistic mass and its associated force law are embedded in the 
physics of a vibrating string, which also (with many additional complications)
 is the physics of the actual fields giving rise to ``particles'' in nature. 
 
We should recognize that Eqs.~(\ref{E}) and \eqref{P} give a rather oversimplified description
of a real string; indeed,
strictly transverse mechanical waves cannot carry longitudinal momentum.\cite{peskin} 
Eqs.~(\ref{E}) and \eqref{P} apply to small-amplitude
oscillations where the string is assumed not to be stretched by the wave
(in which case the wave cannot be strictly transverse).  
The momentum of a field as defined by a formula like Eq.~(\ref{P}) really represents 
energy flux, and it only becomes connected to the velocity of an object 
through the construction of wave packets. 
 
We note for future reference that factors of $\omega$ in wave packet expressions
 are directly proportional to relativistic $\gamma$ factors, as
seen from Eq.~(\ref{omegagamma}). We note also that the results extend 
to packets moving in two or three dimensions by simply replacing $k$ with
a vector $\vec{k}$. For two dimensions the model can still be visualized 
reasonably easily as a vibrating sheet.


\subsection{Relativistic mass in time dilation}

The relativistic mass effect creates a second important mechanism contributing
to time dilation. As an object
accelerates in one direction, the relativistic mass increase of 
its subcomponents results in a slowing of transverse motions  
within the object.
This phenomenon can be understood directly in terms of wave packets. The  
velocity of a wave packet is its group velocity, Eq.~(\ref{gv}), which depends on the overall frequency.
But the frequency measures overall energy, Eq.~(\ref{Ebar2}), and hence is changed by an applied force. 
Acceleration of the wave
packet in one direction increases its frequency and this reduces the group
velocity of the packet in transverse directions. The different components of 
velocity in a wave packet are thus interrelated in a way that would seem
quite unintuitive for a pure point particle. 

A simple example is an orbiting particle  
accelerated slowly perpendicular to the orbital plane (other orbital orientations 
become very complex, hence Bell's original suggestion to study them through numerical simulation).\cite{bell}
There is no tangential force, hence the orbital momentum $P_\perp$ doesn't
change, but since the effective mass does increase one finds that the orbital speed is reduced:
\begin{align}
\label{tangential}
0 &= {d\over dt}(P_{\perp}) = m{d\over dt}(\gamma v_{\perp}),
\end{align}
implying $\gamma v_{\perp} = \mbox{\rm const.}$, or
\begin{align}
\label{tangential3}
v_{\perp} \propto {1\over\gamma}.
\end{align}
Assuming that the orbital radius (or shape, if not circular) stays the same,\cite{angmom}
this result implies that the orbital period increases by the time dilation factor $\gamma$.

This calculation was slightly oversimplified since the particle's $\gamma$-factor 
differs from that of the overall system; however, taking this into account
 leads to the same result.\cite{waverel}
 Also we note that the slower orbit implies a reduced
centripetal force, so the field providing the central force needs to behave accordingly. 
This is not trivial, e.g. a transverse electric field
actually grows with velocity (cf. Fig.~\ref{fig:movingcharge} below), but then a magnetic field also emerges 
whose Lorentz force more than offsets it.

A similar case is the massive analog of the light clock, namely
a``bouncing ball'' clock in which a ball bounces between two plates (with bounce direction
again oriented transverse to the motion of the ship that carries the clock). 
As the ship accelerates perpendicular to the bounce directions, 
the same acceleration must be applied to the ball to keep it bouncing between the two plates. The clock's
caretakers on board the ship must do this without applying any force parallel to 
bounce directions,
because that would invalidate the system's timekeeping function even as seen by themselves. 
Hence they will maintain the clock through small impulses perpendicular to its bounce 
direction, leading to the same slowing effect seen with the accelerated orbit. 

The frequency-changing mechanisms described here and in Sec.~\ref{WPTD} above will affect almost every
type of clock, but they are certainly not an exhaustive list. Most clocks
will also be affected by the shape changes described in Sec.~\ref{WPR} above and Sec.~\ref{LCDPF} below; 
some clocks will also be affected by changes in macroscopic field values, e.g., the electric
and magnetic fields within an LC circuit.


\subsection{Length contraction and deformation of potential fields}
\label{LCDPF}

\begin{figure}[h!]
\includegraphics[width=8.5cm]{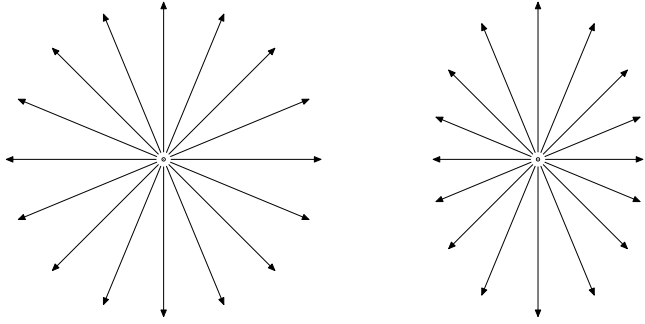}
\caption{Electric field of stationary charge (left) and moving charge (right). Motion is to the right, with $v=0.5c$. Field lines 
indicate field strength on a circle about the charge center, as measured in a stationary frame. }
\label{fig:movingcharge}
\end{figure}

Length contraction---more correctly, shape deformation---first occurred 
to FitzGerald upon seeing Heaviside's solution
showing the deformed electric field about a moving charged particle (see 
Fig.~\ref{fig:movingcharge}).\cite{brown3,heaviside,dmitriyev}  
This famous result  showed that the moving electric field 
becomes ``compressed'' along the direction
of motion, which certainly suggests that any objects constructed using such forces
should undergo at least {\it some} shape change when moving.

Rather than reproduce this computation we note a qualitative way to 
understand why such changes are inevitable. The field of a moving particle 
establishes itself through the emission of electromagnetic waves during acceleration. 
These waves are Doppler-shifted like any other, having a spatial pattern that is asymmetrical
about the charge center. The asymmetry in wave pattern then results in an asymmetrical final field.  
For a scalar field this is the only effect of steady motion, but for the electromagnetic field
a magnetic field is also produced. 

To go further and show that the altered fields actually lead to contraction is worthwhile \cite{contraction}
but we focus here on the qualitative lesson that {\it some}
shape change is inevitable in a field theory. Indeed we assert, with Harvey Brown, that ``shape deformation produced 
by motion is far from the proverbial
riddle wrapped in a mystery inside an enigma.''\cite{brown}

Another way to see the role of internal fields in length contraction is to think of 
how the shape of an object changes as it undergoes acceleration. For concreteness
we imagine a barbell with two identical weights connected by a rod, being accelerated
in the direction along the rod. As it accelerates it also contracts, so
the distance between the weights decreases. Hence the acceleration of the two weights
is not identical; the rear weight accelerates slightly faster than the 
forward one. This in turn implies that the two weights feel slightly different
forces during the acceleration. What is the origin of this force difference? It can
only arise from the changing intermolecular forces within the connecting rod, which
occurs due to the mechanism of Fig.~\ref{fig:movingcharge}.\cite{nikolic}


\subsection{Back-reaction and mass}

Mass/energy equivalence implies that the electric field within a capacitor
adds to its inertia (makes the capacitor harder to accelerate), but how does 
this come about? If one applies
force to the capacitor, the force acts on the atoms forming the plates and housing, not on the 
electric field, so how does the electric field also contribute to the inertia?
 Likewise when an atom emits
a photon and one electron drops to a lower energy level, the atom must become lighter and hence 
easier to accelerate.  Part of this change is due to the reduced electrical field
energy inside the atom, but how does this changed field translate into reduced 
difficulty accelerating the atom?\cite{caveat}

The answer is back-reaction, the process by which a field acts back on its source
to (usually) resist the acceleration of the source. When the relevant field
is electromagnetic, this often amounts to ordinary self-induction. 
It is worth noting that almost all of the mass of everyday matter
actually arises through back-reaction, as manifested in the strong nuclear field.\cite{wilczek}

Back-reaction provides a good illustration of the strengths and weaknesses of both dynamic
and symmetry viewpoints. Using the mass-energy formula immediately gives the
mass contributed by the electric field of a capacitor to be
\begin{align}
\label{cap2}
\Delta M = \Delta E/c^2,
\end{align}
where $\Delta E$ is the standard energy of the electric field between the plates. 
This, however, provides no understanding of how the field actually contributes to the inertia. 

Viewing it mechanistically one sees that self-induction provides at least part of the answer, because  
accelerating the charges on the plates induces a changing $B$ field which in turn
creates an $E$ field that acts back on the charges to resist the acceleration. 
Students can easily calculate this for simple cases such as a parallel-plate capacitor 
or uniformly charged sphere, but a small problem appears: the results don't match  
 the relativistic prediction. Indeed the dynamically computed inertia not only disagrees
but (in the case of a capacitor) depends strongly on the direction of acceleration.\cite{masspaper,kirk}

The problem is that one must also include the fields inside the material of the plates, since it is
these fields that contact the charges directly and exert the back-reaction. One must also then 
consider the motion of the charges inside the material, including those bound within atoms. 
Attempting to do this leads to the even more daunting problem of infinite self-fields of the particles. 
Ultimately one cannot complete the dynamical calculation except using fully renormalized quantum
electrodynamics. 
Nevertheless the naive calculations 
do provide valuable insight into the inertia of field energy; indeed this is how it was 
first discovered, by J.J. Thomson in 1881.\cite{thomp}


\subsection{Time dilation of particle decays}

Particle decay lifetimes are the most commonly observed manifestations of time dilation, 
and the essence of their mechanism can be captured using $\mbox{\rm string}+\mbox{\rm spring}$ models. 
In fact the crucial factor is
 already visible in the driven harmonic oscillator
\begin{align}
\label{dsho}
\ddot{x} + \omega^2 x = F. 
\end{align}

We consider a resonant driving force that turns on at $t=0$; namely,
 $F = \theta(t)\cos{\omega t}$, where $\theta(t)$ is the step function.
Using this driving force, Eq.~\eqref{dsho} has solution
$x = (t/2\omega)\theta(t)\sin{\omega t}$,
showing that the rate of amplitude increase is damped by a factor $1/\omega$. The
same factor is seen more generally in the Green function\cite{mathur}
\begin{align}
\label{shogreen}
G(t-t^\prime) = {1\over\omega}\theta(t-t^\prime)\sin{\omega(t-t^\prime)}. 
\end{align}

The $1/\omega$ suppression carries over to particle decays because, as noted above Eq.~(\ref{Nomega1}),
 the particle fields can be viewed as collections of harmonic oscillators, one for each $\vec{k}$. 
Decays occur when one field drives one or more other fields at resonance, 
and the factor of $1/\omega$ contains
the relativistic dilation factor for moving wavepackets, as noted 
at the end of Sec.~\ref{WPGVRM} above.
(By resonance here is meant that both $k$ and $\omega$ should satisfy Eq.~\eqref{omegak} for the field being driven.)

The most tractable example is not a true decay but rather oscillation between 
two particles having the same mass. This can be modeled by taking two parallel,
identical $\mbox{\rm string}+\mbox{\rm spring}$
systems [Eq.~\eqref{m>0}] and attaching them to each other by additional springs running between them. 
We start with a wave packet only on one of the strings, moving at its group velocity
 $k/\omega$. The packet will then oscillate between the strings and the oscillation frequency will be
``time dilated,'' i.e., it will become slower for faster-moving packets. 

Letting $X$ and $Y$ be the displacements
of the two strings, the springs running between them create a Hooke's law interaction energy
${1\over 2}g(X-Y)^2$. Expanding this, we find that the $X^2$ and $Y^2$ terms just shift the 
mass $m^2$ of each string, leaving the effective interaction energy $-gXY$. 
The coupled equations of motion then take the form (using the shifted value of $m^2$)
\begin{align}
\label{twostringA}
\ddot{X} - X^{\prime\prime} + m^2 X = gY \\
\ddot{Y} - Y^{\prime\prime} + m^2 Y = gX.
\label{twostringB}
\end{align}

A basis of solutions is found by taking $X$ and $Y$ exactly in or out of phase: $X = \pm Y$. 
The out-of-phase modes stretch the springs connecting the two strings, and hence
have higher oscillation frequencies than the in-phase modes; one finds  
\begin{align}
\label{OmegaA}
\Omega_{\pm} &= \sqrt{k^2 + m^2 \mp g} \\
&\approx \omega \mp {g\over 2\omega},
\label{OmegaB}
\end{align}
where the last line is the approximation to first order in $g/\omega$.

A packet that starts out only on the $X$ string can be built by combining
identical packets made with the in-phase and out-of-phase modes. The packets initially
cancel on the $Y$ string, but because they have slightly
different angular velocities $\Omega_\pm$ the initial ``particle'' 
will oscillate between the two strings with angular velocity
equal to the difference: 
\begin{align}
\label{beat}
\omega_{\rm osc} &= \Omega_- - \Omega_+ = {g\over \omega}= {g\over m\gamma}.
\end{align}
The last line uses Eq.~(\ref{omegagamma}) and shows the time 
dilation effect: faster-moving packets 
oscillate more slowly. [The two packets will also separate over time due to their
differing group velocities, however, this effect is of order $O({g/{\omega^2}})$.]
True multiparticle decays can also be understood along these lines but the analysis is more involved.\cite{waverel}


\subsection{Simultaneity}

The relativity of simultaneity is one of the more persistently confusing pieces of the relativity
puzzle, owing perhaps to its nonlocal character, connecting observations 
made at separated locations. Here we give a mechanistic account 
that builds on mechanisms already shown.   

The model is again the transversely-oriented light clock, but this time we consider
two of them. The two clocks start out together at the back end of a moving spaceship,
and they are synchronized. Due to their close proximity they are seen to be synchronized
both by observers onboard the ship and also by external ``stationary'' observers (who we assume,
as usual, to have some way to view the bouncing beams inside the clocks).

Now a scientist onboard the ship carries one of the clocks to the front of the ship. This is
done very slowly in order to avoid disrupting the clock's function. After this ``slow transport''
is complete the onboard observers possess two separated clocks which they can presumably
consider to be synchronized.\cite{simconv} 

\begin{figure}[h!]
\includegraphics[width=8.5cm]{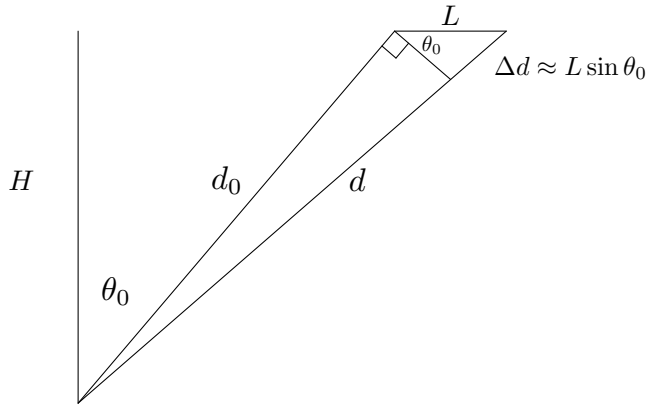}
\caption{Transported light clock.}
\label{fig:lctrans}
\end{figure}

This, however, is not how it appears to the stationary observer, as can be
seen by geometrical analysis similar to that of Fig.~\ref{fig:lightclock}. 
We note first that the mirror reflections have
no effect on the calculation, or equivalently one may give the light clock a height $H$ 
such that it completes exactly one upward bounce during the time taken
to carry it the distance $L$. The situation as seen by the stationary observer is then 
as shown in  Fig.~\ref{fig:lctrans}.

The line tilted at angle $\theta_{0}$ shows the light pulse of a clock aboard the ship
that stays in the same place (is not carried). It completes one vertical pulse of height $H$, traversing a
distance $d_{0}$ as seen from the stationary frame. The line with additional tilt 
 shows the light pulse of the carried clock, which
begins at the same location as the non-carried clock, but covers an additional horizontal distance
$L$, as measured in the stationary frame, and traverses total distance $d$, also as
seen the stationary frame. 

The slow-transport limit is then the limit $H\to\infty$ with $\theta_{0}$ and 
$L$ held constant, and the question is whether the difference in pulse
times goes to zero, or equivalently whether the extra distance $\Delta d\equiv d-d_{0}$ goes to zero.
From the figure one sees that the limiting value is in fact $\Delta d = L\sin\theta_{0}$,
so it approaches zero only when the ship's speed is also zero. The carried clock gets out
of synch with the non-carried clock, as seen by the external observer, no matter how slowly
it is carried. 

The virtue of the slow-transport derivation is that it shows that 
any mechanism causing motion-dependent rate
change also creates motion-dependent synchronization differences. If a given
clock has rate change factor $f(v)$ when moving at speed $v$, then one shows easily 
that slow-carrying in the same direction as the base motion leads
to a synchronization difference $f^\prime(v)L$, where $L$ is the carry distance seen by the 
stationary observer.  One might have thought that
the extra effect would vanish for a clock carried extremely slowly, but this expectation
fails because the slower the clock is carried the more time there is for the
rate difference to accumulate.\cite{edding} 

Comparing this derivation to the more customary ``Einstein train'' thought experiment, 
one sees that {\it for the light clock} the mechanisms are the same.\cite{train}
In both cases the cause of the synchronization discrepancy 
is source-independence of wave speed. 
However, the clock-carrying derivation also extends to other types of clocks whose rate variation
arises from different mechanisms, and 
it also makes sense within theories that have no massless fields at all available for signaling.  
For these reasons we feel that it
captures the underlying mechanism of simultaneity discrepancies between observers, and deserves greater 
emphasis.


\subsection{Cosmic speed limit}

One of the most common questions asked about relativity is why nothing can exceed the speed of light. 
It is difficult to answer this question in a concise and satisfying way. 

One standard answer is that superluminal travel combined with Lorentz invariance 
implies time travel, and hence is paradoxical.  However, this reasoning is quite formal and one 
would hope for a more physical understanding.
A second answer is that the relativistic mass effect makes it impossible to accelerate 
objects to light speed, let alone beyond. This explanation is more physical but still begs the question of why 
 mass acts this way,
and also does not address massless objects.  

If one accepts the premise that everything is ``made from waves'' then 
one can give a more elementary answer. Waves simply cannot be sped up by applied 
forces. Attempting to push on a wave, which one can
easily try at a beach or pool, doesn't make the wave go any faster but only creates 
more waves. Speaking more forcefully does not create faster sound waves but
only louder ones.  Waves can be
slowed down by obstacles that hinder their motion, such as the springs studied in 
Sec.~\ref{WPGVRM} above, but they cannot be sped up. 
In a universe constructed from fields and waves a cosmic speed limit is inevitable by the very
nature of wave motion.

\section{\label{sec:frames}Relating Mechanism and Symmetry}
 
The mechanisms shown above make it clear that in a wave-based world the behavior of moving
objects cannot follow the expectations of Newtonian physics. Moving objects will generically
change shape, while processes within moving objects will not occur at the same rate as when stationary.  

These changes affect all objects and processes, including those used for measurement. This means
that observers in different states of motion will generically measure different values for almost
every quantity,\cite{perspectival} which could produce an extremely complicated situation.  Indeed, 
behaviors could be so complex that neither distance nor time, nor any other customary physical quantities
 can even be meaningfully defined (observers probably could not
 exist under these conditions either). Such generic, very complex field theories
still formally possess one time and three space coordinates, but
it could well be impossible to relate those coordinates to usable operational 
measurements made within those systems.\cite{complex}  

Hence the generic field theory, although still exhibiting relativity-like mechanisms
arising from wave behavior, 
is not likely to be physically interesting. What is needed is a subset of these theories having 
some degree of regularity, say, enough for life to evolve. 
If one began with knowledge of field theory but not of Lorentz invariance one might 
well have looked for theories in which 
the motion-dependent effects are organized in such
a way that arbitrary movements (e.g., orbital or galactic motion) are not fatally disruptive. In
view of the variety and pervasiveness of the mechanisms described above, this is no small order.

Looked at this way it is really rather surprising that there does exist a class of field 
theories in which the effects are beautifully organized and
tuned in exactly such a way that not only can observers exist, but moving observers cannot even
tell they are moving. These are, of course, the Lorentz-invariant field theories. In this very restrictive
class of theories one has concise and operationally meaningful definitions of time, distance, mass, energy, 
and momentum, and their relationships are captured in the elegant formalisms of Minkowski spacetime
and relativistic kinematics. 

Hence the relationship between mechanism and symmetry has something of a chicken-and-egg character. 
The Lorentz symmetry can (apparently) not be realized without the
wave- and field-based mechanisms described above, and yet a generic universe built upon these mechanisms
would likely be barren and uninteresting without the symmetry.\cite{nonlorentz,simconv2}

\section{\label{sec:conc}Conclusion}

Teaching time is obviously limited and one may question whether it is 
productive to spend time on discussions that mainly add qualitative 
understanding. 
We feel that in the case of relativity it should be seriously considered in view of the absolute centrality
of the concepts involved. There is no other part of the curriculum that
deals primarily with the core concepts of distance, time, energy, mass, and measurement
that permeate all physical thinking.

The surprising ease of deriving the Lorentz transformation can act, paradoxically, 
as a barrier to full understanding. Abstract explanations based solely 
on postulates or symmetry
hide the true complexity of the underlying processes and do not provide a complete foundation for
reasoning about the fundamental concepts involved. 
Many students are left with lingering sensations of circularity or incompleteness in the 
derivations as well as serious uncertainties about what the theory covers, what the alternatives to 
Lorentz invariance are, and how effects such as length contraction
relate to more familiar physical effects. Consideration of the mechanisms
of relativity unifies it more closely to other areas of physics and should help to
forestall these sorts of confusions.

\section{Acknowledgments}
The author gratefully acknowledges helpful communications with and suggestions from Dean Welch,
 Larry Hoffman, Michael Lennek, Francis Everitt, Bryan Galdrikian, Shirley Pepke and Kirk McDonald. 


\begin{thebibliography}{5}
\bibitem{dynamic} These authors also proposed the label ``dynamical'' for this viewpoint, 
which we have adopted (Ref.~\onlinecite{brown5}).
\bibitem{coupling} Another possibility for the coupling involves one time derivative, e.g.
$\phi(x)\xi\dot{\xi}$, which is similar to the coupling of an electric potential. 
A coupling with two time derivatives would be analogous to a gravitational potential.
\bibitem{packetnorm} Derivatives convert the cosines of Eq. (5) into sines, however,
 the packet built on sines is approximately equal to the packet built on cosines and shifted by $\Delta x = -{\pi/2k}$.
\bibitem{angmom} This can be shown using conservation of angular momentum, but this is of 
course one aspect of Lorentz invariance, since the Lorentz group contains rotations.
\bibitem{contraction} Ideas for calculating explicit shape change can be
found in Ref.~\onlinecite{miller} and Ref.~\onlinecite{bell} (the latter proposes numeric calculations of orbiting systems 
undergoing acceleration). Calculations to lowest order in $v/c$ are also feasible.
\bibitem{caveat} The other part of the mass change is actually an increase, because the electron in a lower
energy state moves faster and hence has a larger relativistic mass. The virial theorem
guarantees that this increase is more than offset by
the decrease in field energy.
\bibitem{complex} The fact that formal quantities, especially the metric, may not have their
 expected operational meaning within the theory, has also been discussed by Brown (Ref.~\onlinecite{brown2})
 and Brown and Pooley (Ref.~\onlinecite{brown4}).
\bibitem{simconv2} This helps understand the question of conventionality of the definition of
simultaneity (see, e.g., Ref.~\onlinecite{brown}). If Lorentz invariance holds then there is a preferred definition of
simultaneity and little reason to consider any other; otherwise one will need to consider conventions, but probably no really ideal convention will exist. In no case can a field/wave world be converted back to Newtonian behavior by means of conventions.
\end{thebibliography}
\end{document}